**Title**

A Global Map of Suitability for Coastal *Vibrio cholerae* Under Current and Future Climate Conditions

**Authors**

Luis E. Escobar[1,2*], Sadie J. Ryan[1,3,4], Anna M. Stewart-Ibarra[1,5], Julia L. Finkelstein[6], Christine A. King[1,7,8], Huijie Qiao[9], Mark E. Polhemus[1]

**Affiliations**

[1]Center for Global Health and Translational Science, State University of New York (SUNY) Upstate Medical University, Syracuse, NY, USA

[2]Facultad de Ecología, Universidad Andres Bello, Santiago, Chile

[3]Department of Geography and Emerging Pathogens Institute, University of Florida, Gainesville, FL, USA

[4]School of Life Sciences, College of Agriculture, Engineering, and Science, University of KwaZulu-Natal, Durban, South Africa

[5]Department of Medicine, SUNY Upstate Medical University, Syracuse, NY, USA

[6]Division of Nutritional Sciences, Cornell University, Ithaca, NY, USA

[7]Department of Microbiology and Immunology, SUNY Upstate Medical University, Syracuse, NY

[8]Laboratorio de Biomedicina, Escuela Superior Politécnica del Litoral, Guayaquil, Ecuador

[9]Key Laboratory of Animal Ecology and Conservation Biology, Institute of Zoology, Chinese Academy of Sciences, Beijing, China

**\*Corresponding Author:**

Luis E. Escobar, ecoguate2003@gmail.com, Phone: 315-464-8153, Fax: 315-464-4417

**Short Title**

Coastal *Vibrio cholerae* and Climate Change

**Abstract**

*Vibrio cholerae* is a globally distributed water-borne pathogen that causes severe diarrheal disease and mortality, with current outbreaks as part of the seventh pandemic. Further understanding of the role of environmental factors in potential pathogen distribution and corresponding *V. cholerae* disease transmission over time and space is urgently needed to target surveillance of cholera and other climate and water-sensitive diseases. We used an ecological niche model (ENM) to identify environmental variables associated with *V. cholerae* presence in marine environments, to project a global model of *V. cholerae* distribution in ocean waters under current and future climate scenarios. We generated an ENM using published reports of *V. cholerae* in seawater and freely available remotely sensed imagery. Models indicated that factors associated with *V. cholerae* presence included chlorophyll-a, pH, and sea surface temperature (SST), with chlorophyll-a demonstrating the greatest explanatory power from variables selected for model calibration. We identified specific geographic areas for potential *V. cholerae* distribution. Coastal Bangladesh, where cholera is endemic, was found to be environmentally similar to coastal areas in Latin America. In a conservative climate change scenario, we observed a predicted increase in areas with environmental conditions suitable for *V. cholerae*. Findings highlight the potential for vulnerability maps to inform cholera surveillance, early warning systems, and disease prevention and control.

1. Introduction

Cholera, a water-borne disease caused by *Vibrio cholerae*, remains a severe threat to public health and development globally (WHO, 2013). According to the World Health Organization (WHO), the burden of cholera is more than 100 times greater than current estimates (WHO, 2013; Zuckerman et al., 2007), with 120,000 deaths and three to five million cases per year worldwide. Previous studies suggest that cholera outbreaks are associated with oceanographic variables such as water temperature, pH, salinity, and phytoplankton blooms, indicating a potential to predict disease outbreaks (Jutla et al., 2013, 2010). A 2010 analysis of global cholera pandemics found that cholera outbreaks originate in coastal regions, often during flood events, before spreading inland (Jutla et al., 2010).

Based on the physiological requirements for growth and phylogenetic information of the species, it has been suggested that the original ecosystem of *V. cholerae* is the sea (Colwell, 2004). However, *V. cholerae* persists environmentally in riverine, estuarine, and coastal waters around the world (Lipp et al., 2002). It is endemic in the Bay of Bengal (Bangladesh and India) and along coastal areas in Latin America (Lipp et al., 2002; Mutreja et al., 2011). Cholera is caused by *Vibrio cholerae* a gram-negative bacterium, classified in at least 200 serogroups, with toxogenic serogroups O1 and O139 often implicated in human epidemics (Harris et al., 2012). It has been a reportable disease since the 19$^{th}$ century in the United States (Sack et al., 2004), and remains an internationally notifiable though neglected disease (Ryan, 2011). It was one of the first diseases validated by Koch's postulates (Glass et al., 1992), and the 1854 cholera outbreak in Soho, London, is the subject of the classic John Snow Broad Street Pump spatial epidemiological case study (Koch and Denike, 2009). Despite the accumulated knowledge of this classic water-borne disease, the current seventh pandemic continues (Sack et al., 2004), and novel emerging serogroups may be sources of a new eighth pandemic (Kaper et al., 1995). Cholera is a major threat to the health of vulnerable populations, generating opportunistic epidemics after natural disasters in human settlements with precarious conditions, including inadequate access to potable water and sanitation, during periods of both very low or high precipitation (Pascual et al., 2002).

Epidemics of *V. cholerae* follow coastlines (Colwell, 1996) and the pathogen can be transmitted by a wide range of marine organisms, including zooplankton, aquatic plants, shellfish, and fish (Vezzulli et al., 2010). There is also evidence that *V. cholerae* can be mechanically translocated by arthropods, protozoa, aquatic birds, and marine mammals (Vezzulli et al., 2010). Additionally, *V. cholerae* aggregate and form biofilms that allow adherence to both abiotic and biotic surfaces, including plankton. *V. cholerae* biofilms, while non-culturable, are more resistant to adverse environmental conditions compared to single cells, and infectious when consumed by the host (Alam et al., 2007). Thus, biofilms are a key component in *V. cholerae* ecology and epidemic seasonality (Alam et al., 2007); these features may make *V. cholerae* resistant during anthropogenic water translocation. McCarthy and Khambaty found that ballast water from cargo ships contain *V. cholerae*, and therefore represent a means to facilitate worldwide translocation (McCarthyl and Khambaty, 1994). *V. cholerae* translocation to non-endemic areas can have a catastrophic impact on the local human population. A recent *V. cholerae* translocation contributed to the 2010 epidemic in Haiti, with over 3,071 cases per 10,000 population (Gaudart et al., 2013). The diversity of transmission mechanisms means that cholera outbreaks can also result from contaminated seafood consumption, even in developed countries with high-quality water supply (Blake et al., 1980).

Marine environments are a key component of *V. cholerae* ecology, and current international trade of ocean-derived materials presents a constant threat for *V. cholerae* translocation. Climate is a critical component of the *V. cholerae* ecological niche. *V. cholerae* detection increases with environmental water temperatures (Lipp et al., 2002), likely reflecting an increase in bacterial abundance and concentration (Confalonieri et al., 2007). Explorations of cholera epidemiology using remote sensing data, revealed an association of sea surface temperature (SST) with cholera occurrence in Bangladesh (Colwell, 1996). Lipp et al. (2002), based on a review of quantitative evidence, also suggested that increasing seawater temperatures projected under future climate change may drive *V. cholerae* to emerge in new regions of the world (Lipp et al., 2002). Trends in historic SSTs point to increasing maximum SSTs in the past few decades (Supplementary Material S1), indicative of this climate forcing. Additionally, and intrinsic to future *V. cholerae* ecology, climate change is predicted to increase both water temperature and flood events, likely raising the risk of cholera infection and transmission in vulnerable coastal populations (Confalonieri et al., 2007).

Understanding the factors that influenced the emergence of past cholera epidemics may help inform monitoring and prevention of future outbreaks (Lipp et al., 2002). The current cholera pandemic emerged in the Americas in 1991 from rural areas of coastal Peru and Ecuador (Swerdlow et al., 1992). Determining the biophysical conditions that promote the growth and persistence of *V. cholerae* in environmental waters is vital to predicting vulnerability and risk, development of early warning systems, and disease prevention and control (Kuhn et al., 2005). Furthermore, identifying vulnerable areas and high-risk populations in the face of future climate changes may help inform disease prevention and control (Confalonieri et al., 2007).

The objective of this study was to identify environmental variables associated with *V. cholerae* presence and to generate a global map of *V. cholerae* risk, testing the hypothesis that the bacterium can be predicted in the ocean with statistical significance, using a series of oceanographic variables. We determined the environmental variables associated with *V. cholerae* occurrence in seawaters and established potential *V. cholerae* distribution maps in coastal areas worldwide under current and future climate scenarios. We conducted a structured literature review to identify and georeference sites of laboratory-confirmed *V. cholerae* presence in seawaters. Using environmental data from satellite imagery, we developed an ecological niche model (ENM) to predict global *V. cholerae* presence and identify high-risk areas under current and future climate scenarios.

2. Methods

*2.1 Sites of V. cholerae presence*

In order to identify prior studies of the geographic distribution of *V. cholerae* in ocean waters, we conducted a structured literature search using Web of Science (WOS) electronic databases from January 1950 to April, 2014. We used the key words *Vibrio cholerae* and geographic distribution*, spatial distribution*, spatial analysis*, spatial epidemiology*, biogeography, map, and mapping. Initial inclusion criteria for this review were the availability of geographic location of *V. cholerae* reports (e.g., latitude and longitude coordinates or detailed identification of geographic location of reports), regardless of date. Full-text articles were extracted and reviewed, and the following inclusion criteria were applied: studies reporting *Vibrio cholerae* collected in seawater (environmental samples) and confirmation by laboratory analysis.

*2.2 Study areas*

The study area location and extent selected for inclusion in an ecological niche model have a strong influence on model calibration (including weights given to particular environmental variables) when modeling the potential presence of an organism and in model performance (Barve et al., 2011; Peterson and Nakazawa, 2007). We determined the site location of *V. cholerae* reports based on data extracted in our literature review, including areas in the Northern Hemisphere, the region off the coast of Ecuador, and the southern Hemisphere (Results). In order to reduce the potential impact of the study area size on the *V. cholerae* model, we buffered report sites by 100 kilometers. This buffer radius was selected based on the distance between *V. cholerae*-positive sampling points found in literature (Binsztein et al., 2004). The ENMs were calibrated and evaluated on these buffered zones, and transferred onto global scale surfaces (at ~9 km resolution), using current and future climate scenarios. These models are described in the following section.

*2.3 Environmental variables*

We used 12 environmental variables to build the models, based on previous ENMs of shallow-water organisms in marine environments (Tyberghein et al., 2012). These annual variables included minimum, mean, and maximum SST ($^{o}$C); maximum photosynthetically available radiation (Einstein/m$^2$/day), mean salinity, mean pH, mean dissolved oxygen (ml/L), mean nitrate and phosphate (µmol/L); and minimum, mean, and maximum chlorophyll-a (mg/m$^3$) at approximately 9-kilometer resolution. The variables were obtained from Bio-ORACLE (Tyberghein et al., 2012), a marine environment database derived from monthly satellite imagery collected by sensors Aqua and Terra MODIS and SeaWiFS between 2005 and 2010 (http://www.oracle.ugent.be/).

These data were used for exploratory analyses for the ENM, interpreting Maxent's logistic output as a suitability index ranging from 0 for low suitability to 1 for high suitability to generate a risk map under current climate conditions to describe *V. cholerae* suitability according to the environmental variables used (Merow et al., 2013). We used a moderately conservative climate change scenario (B1) (Tyberghein et al., 2012), from the World Climate Research Programme Coupled Model Intercomparison Project (Jueterbock et al., 2013), to predict potential future *V. cholerae* distribution. Using the B1 scenario, the ENM created under current conditions was transferred and projected to the future scenario using mean salinity and mean, minimum, and maximum SST; values were transferred and projected for the year 2100 at 9-kilometer resolution.

*2.4 Statistical analyses*

We conducted a Principal Components Analysis (PCA) to identify highly correlated environmental variables and reduce dimensionality of the model. PCA is a multivariate analysis that generates a correlation matrix to identify the most associated variables *via* correlation coefficients, and is commonly used in satellite imagery analysis in ecology (Horning et al., 2010). The original number of environmental variables (12) was reduced by removing highly correlated variables (correlation coefficient >0.55, see Supplementary Material S2) to avoid collinearity and model overfitting (Peterson et al., 2011), and the remaining four original, uncorrelated variables were used in model calibration (Peterson et al., 2011).

We developed an ENM *via* a presence-background algorithm (Guillera-Arroita et al., 2014; Peterson et al., 2011), using Maxent version 3.3.3.k (Phillips et al., 2006). Maxent is a correlative model based on the maximum entropy principle, that identifies the most uniform probability of presence distribution, subject to environmental constraints and assumptions (Anderson, 2013; Franklin, 2009; Phillips et al., 2006). We created a correlational model, bootstrapped with 500 permutations; we used 20% of the points from sites where *V. cholerae* was reported for model evaluation and the remaining points to calibrate the model. As a final output, we used the median of the permutations on the logistic format with values ranging from zero to one as estimates of relative suitability. In order to transfer the model from the buffered zones to a global extent, we used Maxent with clamping and extrapolation turned off (i.e., no prediction outside the range of environmental conditions in the calibration areas) to avoid the risk of over prediction in non-analogous environments (Anderson, 2013; Guisan et al., 2014).

In the final model calibration, we used a jackknife test to evaluate the model performance for each of the environmental variables (Elith et al., 2011). This enabled us to determine the relative influence of the variables on *V. cholerae* prediction. The calibrated *V. cholerae* ENM was transferred and projected to a global scale under current and future climate scenarios using Maxent's projection tool for model transference and extrapolation (Anderson, 2013). Model transference is defined as prediction in novel areas, considering only those environments available in the calibration area (i.e., no prediction into novel environments), while extrapolation allows prediction into novel environments (Anderson, 2013). We generated a conservative model of strict model transference (i.e., no extrapolation or clamping) to avoid perilous extrapolation (see Owens et al., 2013). For visual comparison, we also explored model extrapolation (i.e., turning on extrapolation and clamping options in Maxent) allowing for the prediction of *V. cholerae* in environments not available in calibration areas (Anderson, 2013).

Environmental conditions in sites with *V. cholerae* were compared in a three-dimensional space built with the first three principal components from the PCA to capture most of the information contained in the environmental variables. Sites with *V. cholerae* present in ocean waters were plotted in the gradient space in the form of minimum-volume ellipsoids (MVEs) generated around points of *V. cholerae* presence (Aelst and Rousseeuw, 2009), to visualize the position and volume of each site in the environmental space. The software NicheA requires a minimum of four points summarizing environmental information to construct the ellipsoids in a three-dimensional scenario to provide volume. In order to quantify similarities between sites, we calculated a Jaccard similarity coefficient, measuring the amount of overlap of the ellipses of each site, defined below (Equation 1):

$$V_0 = \frac{V_1 \cap V_2}{V_1 \cup V_2}$$

(1)

where $V_0$ is the volume of the overlap, $V_1$ is the volume of the first ellipsoid, and $V_2$ is the volume of the second ellipsoid. Analyses were conducted using NicheA (Qiao et al., 2015), an open-source software for ecological niche analysis developed by Qiao et al. (training videos and detailed information on the algorithms used are available at http://nichea.sourceforge.net/). Additional data management and analysis was conducted using R version 3.1.0 (R Core Team, 2012) in the RStudio platform (RStudio, 2014), QGIS (http://www.qgis.org/), and ArcGIS version 10.1 (ESRI, 2012).

In order to assess the predictive ability of the algorithm, variables, and occurrences employed, we used a "quadrant test" of equal sample size (Levine et al., 2007). Briefly, the *V. cholerae* occurrence data set was divided into four quadrants from available occurrence points; two off-diagonal quadrants were used for model calibration and two were used for evaluation, and then the reverse procedure was used to obtain two independent evaluations (Peterson et al., 2011). From the occurrences not used in model calibration, we obtained the evaluation areas and points for a binomial cumulative probability (BCP) test (Peterson et al., 2011), in which the total number of evaluation points were trials, the number of evaluation points predicted correctly were successes, and the proportion of the evaluation area predicted suitable was defined as the probability of success (Anderson et al., 2002). In order to estimate the proportion of suitable area during model evaluation, we generated a binary map using 90% of calibration occurrences as threshold, considering potential error in occurrence geolocation (Peterson et al., 2011). This threshold was developed in ArcGIS to ensure the minimum area that contains 90% of occurrences.

Finally, to distinguish novel environmental conditions under future climate conditions (i.e., environments not existing at present), we used two analytical approaches: the Mobility-Oriented Parity (MOP) and the Multivariate Environmental Similarity Surface (MESS) (Elith et al., 2010; Owens et al., 2013). MOP identifies future environmental conditions not available in present climate conditions, while MESS assesses the level of similarity between climate scenarios. Results allowed us to establish those areas of strict model extrapolation from those areas with environmental conditions present in any site on Earth. This analysis was performed in R with a sample of 10% of the available cell in the ocean variables.

## 3. Results

Our literature search returned 759 articles; of these, we found 15 sites with *V. cholerae* reports in seawaters, including coastal areas of Peru (4), the United States (1), Colombia (4), Venezuela (1), Brazil (1), Argentina (2), Germany (1), and Bangladesh (1) (Binsztein et al., 2004; Grau et al., 2004; Lipp et al., 2003; López et al., 2010; Louis et al., 2003; Martinelli Filho et al., 2010; Mukhopadhyay et al., 1998; Orozco et al., 1996; Pal et al., 2006; Pascual et al., 2000). All sites were georeferenced except Bangladesh, which we geolocated 100 kilometers (117 pixels) off of the Bakerganj coast in the Bay of Bengal (Sack et al., 2003), since we wanted to include this important *V. cholerae*-endemic area, but reported geographic detail was coarse. The 15 sites were used to characterize marine environmental variables that influence potential *V. cholerae* distribution, representing diverse marine environments (Fig. 1).

*V. cholerae* occupied a broad range of environmental conditions, including chlorophyll-a, oxygen, and temperature, and a narrow range of pH (Fig. 2). Results from the PCA indicated correlations among mean and minimum chlorophyll-a values, and maximum and minimum SST. Thus, correlated variables were removed to calibrate the model (Supplementary Material S2). Four variables were included in the final ENM: mean chlorophyll-a, pH, maximum SST, and mean salinity. Environmental variables across the sites of *V. cholerae* presence ranged from -1.53 to 34.46 °C maximum SST, 7.52 to 8.34 pH, 4.14 to 7.46 ml/L dissolved oxygen, 5.36 to 36.72 PSS salinity, and 21.88 to 22.08 mg/m$^3$ chlorophyll-a (Fig. 2). In the final model, the relative variable contribution was 49% for mean chlorophyll-a, 20% for pH, and 16% for maximum SST. In the jackknife analysis, mean chlorophyll-a was the most informative variable explaining *V. cholerae* presence, and the variable that most affected the model performance when omitted. The first three principal components (PCs) explained 97.1% of the variance in the original environmental variables (bands 1, 2, and 3 in Fig. 1). By plotting these three components in a three-dimensional space (Fig. 3) and measuring the ellipsoids' overlap in the environmental space (Table 1), we identified the overlap between the coast of Argentina and Chesapeake Bay in the United States, between the Bay of Bengal and regions of Venezuela and Peru, and between areas of southern Brazil and Venezuela (Jaccard index >0; Table 1). The similarity was not appreciable between other areas in this study (Jaccard index=0). Furthermore, our model evaluations from the quadrant test revealed that the predictions were significantly non-random ($p<0.001$).

Under current climate conditions, we found suitability for *V. cholerae* in coastal regions on all continents (Maxent suitability index range: 0.02-0.94), as shown in the global potential distribution map generated through the ENM (Fig. 4). However, the most suitable areas (>0.49) for potential *V. cholerae* distribution were coastal Peru and Ecuador in South America, Canberra in Australia, the Bay of Biscay and North Sea in Europe, the Gulf of California, the Yellow Sea, and the Gulf of Oman, coastal Mauritania, Senegal, and Western Sahara, and large areas in South-Western Africa (Fig. 4). At a finer scale, additional areas with high *V. cholerae* suitability were found along the coasts of Kenya, Maldives, Bangladesh, Thailand, Vietnam, and Cambodia (Supplementary Material S3). Models of the impact of future climate conditions (under scenario B1) on *V. cholerae* presence demonstrated an increase in potential areas for *V. cholerae* distribution in both model transference and extrapolation (Fig. 4). According to our model, by 2100, there is potential for a latitudinal expansion of suitable areas for *V. cholerae*. Findings were more conservative in the transference model compared to the extrapolative model, which indicated higher levels of suitability even in extreme environments at higher latitudes (Fig. 4).

When extrapolation was allowed, the *V. cholerae* model ranged from -1.87 to 41.83 °C SST and 0 to 45 PSS salinity.

The MESS analysis demonstrated broad areas with similar environmental conditions between current and future climate conditions. MOP analyses revealed sites in the Caribbean and in the South China Sea not available under current climate conditions (Supplementary Material S4). ENM transferred to future climate do not included predictions on environments absent at present (black areas in S4), while ENM with extrapolation identified suitable areas in novel environmental conditions (Fig. 5).

## 4. Discussion

*4.1 Explanatory variables*

Cholera remains a global threat to human health and development. Over the last two centuries since it was first recognized, new virulent strains continue to arise and are expected to continue to evolve and spread (Sack et al., 2004). In this study, we identified macro-ecological patterns of *V. cholerae* in global seawaters. Chlorophyll-a and pH were the most important variables in the *V. cholerae* presence in the model. These variables interact during phytoplankton photosynthesis (Lipp et al., 2002), providing important bounds on fundamental processes for cholera persistence. In fact, remotely detected chlorophyll-a, the main pigment found in phytoplankton (Horning et al., 2010), was the most important explanatory factor in the *V. cholerae* model. A broad range of environmental conditions are suitable for *V. cholerae* (Lipp et al., 2002): according to our model, *V. cholerae* may be present at lower temperatures and higher pH than indicated by the empirical data, as shown in Figure 3 (red line). Previous studies of *V. cholerae* occurrence in field and laboratory analyses demonstrated bacterium tolerance to temperatures from 0.2 to 30 °C, salinity from 2.5 to 30 PSU, and pH from 6.3 to 7.5 (Kaper et al., 1979; Louis et al., 2003; Miller et al., 1984). Allowing model extrapolation generated extreme values for *V. cholerae* presence (i.e., -1.87 to 41.83 $^{o}$C SST and 0 to 45 PSS salinity; Fig. 5). However, tolerance of *V. cholerae* to these values lack physiological support (Owens et al., 2013).

The environmental parameters identified in the *V. cholerae* ENM under current climate conditions were consistent with previous studies conducted in Bangladesh. Constantin de Magny et al., (2008) found an association between high levels of chlorophyll-a in seawater and number of human cholera cases in two cholera endemic areas in the Bay of Bengal, while Emch et al., (2008) found that an increase in ocean chlorophyll was associated with an increase in cholera cases, with a two-month lag. Other studies have examined the associations between seawater variables and in-land reports of cholera cases at a local geographic scale (Colwell, 1996; Constantin de Magny et al., 2008; Jutla et al., 2013; Lobitz et al., 2000). Although these studies provide support for our findings, large-scale, global validation is not yet feasible.

Previous studies of cholera in coastal waters identified important environmental risk factors such as water temperature, algal presence, salinity, zooplankton, and water volume (Constantin de Magny et al., 2008; Huq et al., 2005; Kaper et al., 1979; Pascual et al., 2002; Paz and Broza, 2007). Inconsistent findings regarding the factors that trigger cholera emergence could be due to differences in the spatial (e.g., study area extent, grid size) and temporal (e.g., one measure, seasonal measures) resolution of studies, or differences in the ecological systems (e.g., freshwater *versus* seawaters). For example, Louis et al. (2003) measured surface water temperature *in situ* in an estuarine system and found that temperature was associated with cholera, while Emch et al. (2010) used satellite-derived variables in seawaters and found that surface water temperature was not associated with cholera incidence (Emch et al., 2010; Louis et al., 2003). To our knowledge, the current study is the first to link seawater environments to *V. cholerae* presence in seawater, at a global scale. However, our results are scale-dependent, so we caution against inappropriate extrapolation of our model to different spatial scales (Escobar et al., 2013). Furthermore, the restrictive reduction of correlated variables to reduce model overfitting may limit the ability to identify other important variables that affect the *V. cholerae* model (Supplementary Material S2) (Merow et al., 2013).

*4.2 Potential for spread*

Phylogenetic studies of *V. cholerae* samples globally have identified the Bay of Bengal as the epicenter of the ongoing pandemic (Mutreja et al., 2011). In the current study, we observed similarities in the ecological environment of the Bay of Bengal on the coast of Bangladesh and coastal areas of Latin America (Fig. 4). This supports the hypothesis that mechanical translocation to similarly suitable environments may trigger *V. cholerae* emergence in distant geographic areas (Fig. 2).

Further empirical field research is needed to better understand *V. cholerae* distribution in coastal habitats and to identify drivers for disease emergence. Our *V. cholerae* suitability map (Fig. 4) is currently in use in a field study in coastal Ecuador, to identify patterns of *V. cholerae* presence and environmental covariates, providing fine-scale data for model validation. Additional sites of *V. cholerae* presence may enhance the predictive ability of our model, although the small number of sites (n=15) included in our model spanned a wide geographic distribution, providing diverse environmental information (Fig. 2).

We developed a conservative ENM to define potential *V. cholerae* distribution areas worldwide, based on available data from sampled areas, avoiding predictions in marine environments outside of sampled regions. This approach has been used extensively in invasion biology to reduce predictions in environments outside of those from the calibration areas (Guisan et al., 2014). Consequently, even when predicted areas for potential *V. cholerae* distribution appear to be extensive (Fig. 4), this area may increase if *V. cholerae* is identified in novel marine environments.

*4.3 Climate change*

Projected climate data sets exist for past and future terrestrial climate scenarios (www.worldclim.org), but fewer explorations exist for marine environments. This is due to limited availability of data and the complexities of marine variables (e.g., currents, minerals, pH, and oxygen), which may have dynamic interactions that are challenging to accurately predict under climate change scenarios. Therefore, we restricted the future models to salinity- and temperature-based prediction models, given the high degree of confidence of anticipated alterations in salinity, temperature, carbon dioxide concentration, and available oxygen in the ocean; as a result marine communities will respond to these changes (IPCC, 2014). In the moderate climate change scenario (B1), the potential future geographic distribution of *V. cholerae* increased, compared to the current climate scenario, including latitudinal expansion in *V. cholerae* range and an increment of suitable conditions in open waters (Fig. 5). In the future climate scenario, the transference model identified several areas with suitable conditions for *V. cholerae* growth, including the Pacific coast off of western North America, the Gulf of Saint Lawrence off the coast of Quebec, Canada, the Coral Sea and the Tasman Sea from Fiji to New Zealand, and the Black Sea in Southeastern Europe. These findings provide insight to inform public health planning in high-risk coastal areas. Additionally, in the transference model, there was a shift in the range of *V. cholerae* suitability from low to medium and high latitudes (Fig. 5). Strikingly, this pattern of range shifts parallels that of plankton communities, which have shifted northward in recent decades in response to increasing seawater temperatures (IPCC, 2014). Modeling marine distribution of pathogens under future climate scenarios is a powerful tool to inform effective decision making and prioritization of public health resources (IPCC, 2014). However, using ENMs to anticipate species range response under future climate conditions must be interpreted with caution, considering our limited understanding of physiological response of species to climate change (Merow et al., 2013). For example, in future climate scenarios, when *V. cholerae* model was allowed to extrapolate the prediction under extreme latitudes (i.e., environments not available in the calibration areas), this resulted in ENM overprediction (Fig. 5), as previously reported when ENMs are generated under uncontrolled situations (Owens et al., 2013). However, using model extrapolation (Fig. 5, bottom panel) – instead of model transference – generates dramatic models with a lack of biological realism (Owens et al., 2013). In fact, our MOP analysis of the environments not existent in the ocean at present (Supplementary Material S4), allowed us to identify areas of model extrapolation where suitable conditions are anticipated without biological realism (see black areas in Supplementary Material S4 and discussions in Owens et al. 2013). Modeling species distributions under future climate conditions requires the comparisons of current and future climates to provide robustness for model selection and interpretation. Here, we found that model transference is the most adequate method to forecast species distribution under future marine conditions, as transference avoids prediction in inexistent environmental conditions (Supplementary Material S4).

We based our predictions on laboratory-confirmed reports of the *V. cholerae* species, thus, we considered pathogenic and non-pathogenic strains during model calibration. The tendency of species to maintain phylogenetic niche conservatism (Peterson et al., 1999) provides support for the application of our model to forecast pathogenic and non-pathogenic strains and other vibrios phylogenetically close to *V. cholerae* at the spatial scales considered here (see discusion in Tocchio et al., 2014). Niche conservatism also allows us to assume that the species' niche will be conserved under future climate conditions (Peterson et al., 1999).

*4.4 Future research*

Future studies should be conducted in coastal regions with ongoing cholera outbreaks to validate current global models. Additional field research from other coastal regions with ongoing cholera outbreaks that also have *V. cholerae* present in ocean waters (e.g., Haiti) could provide environmental information to improve model calibration. The current *V. cholerae* model was developed using abiotic parameters; however, interactions with vectors (e.g., copepods) play a crucial role in *V. cholerae* spread and persistence. Future research is needed to investigate the interactions among abiotic and biotic factors that influence *V. cholerae* persistence in coastal waters in order to determine the current and future risk of cholera transmission. Findings from this study confirm that *V. cholerae* is a generalist bacterium, distributed across a broad range of chlorophyll-a and temperature. This ecological plasticity explains, in part, the success of this pathogen in spreading and persisting across a broad range of water features. Furthermore, the global distribution (i.e., from sea to fresh waters) underscores that cholera is not simply a disease of poverty. However, even when we anticipate that environmental and climate factors are predicted to favor *V. cholerae* occurrence in the future, it is clear that economic, social, immunological, and cultural factors directly contribute to cholera risk at local scale (Confalonieri et al., 2007). Importantly, social-ecological variables (vulnerabilities) not considered in this study, such as access to sewers, interact with biophysical conditions (hazards) to influence the probability (risk) of current and future cholera outbreaks. Future studies could be strengthened through the inclusion of social-ecological model constraints (e.g., access to piped water, socioeconomic status) and demographic projections.

*4.5 Final remarks*

Epidemiological studies that include maps of reported disease occurrence without spatial (statistical) analyses of their distribution associated with environmental features my overlook key ecological drivers of transmission (Ruiz-Moreno et al., 2010). Since seafood is a main source of human cholera cases (Vezzulli et al., 2010) and zooplankton play a key role in cholera ecology and putative spread (Huq et al., 1983), future studies should evaluate the role of marine vectors (e.g., zooplankton) in cholera distribution at a global extent. Our ecological niche modeling approach for *V. cholerae* may represent an important tool to identify potential areas for cholera occurrence in coastal environments, and inform public health prioritization, surveillance, and monitoring of coastal and estuarine waters in high-risk areas, and disease prevention and control. The complimentary approaches of ecological modeling and surveillance of abiotic factors are key strategies to target cholera epidemics.

**Acknowledgements**

LEE, SJR, and AMS were supported by the Global Emerging Infectious Disease Surveillance and Response System (GEIS) Grant P0435_14_UN.

**Table 1. Jaccard index comparisons of similarity.** Similarities were measured on environmental conditions between two sets of minimum-volume ellipsoids representing *Vibrio cholerae* presence by region.

|  | Argentina | Bangladesh | Brazil | United States | Colombia | Germany | Peru | Venezuela |
|---|---|---|---|---|---|---|---|---|
| Argentina |  | 0 | 0 | **0.027** | 0 | 0 | **0.033** | 0 |
| Bangladesh |  |  | 0 | 0 | **0.04** | 0 | 0 | **0.115** |
| Brazil |  |  |  | 0 | 0 | 0 | 0 | **0.007** |
| United States |  |  |  |  | 0 | 0 | 0 | 0 |
| Colombia |  |  |  |  |  | 0 | 0 | 0 |
| Germany |  |  |  |  |  |  | 0 | 0 |
| Peru |  |  |  |  |  |  |  | 0 |

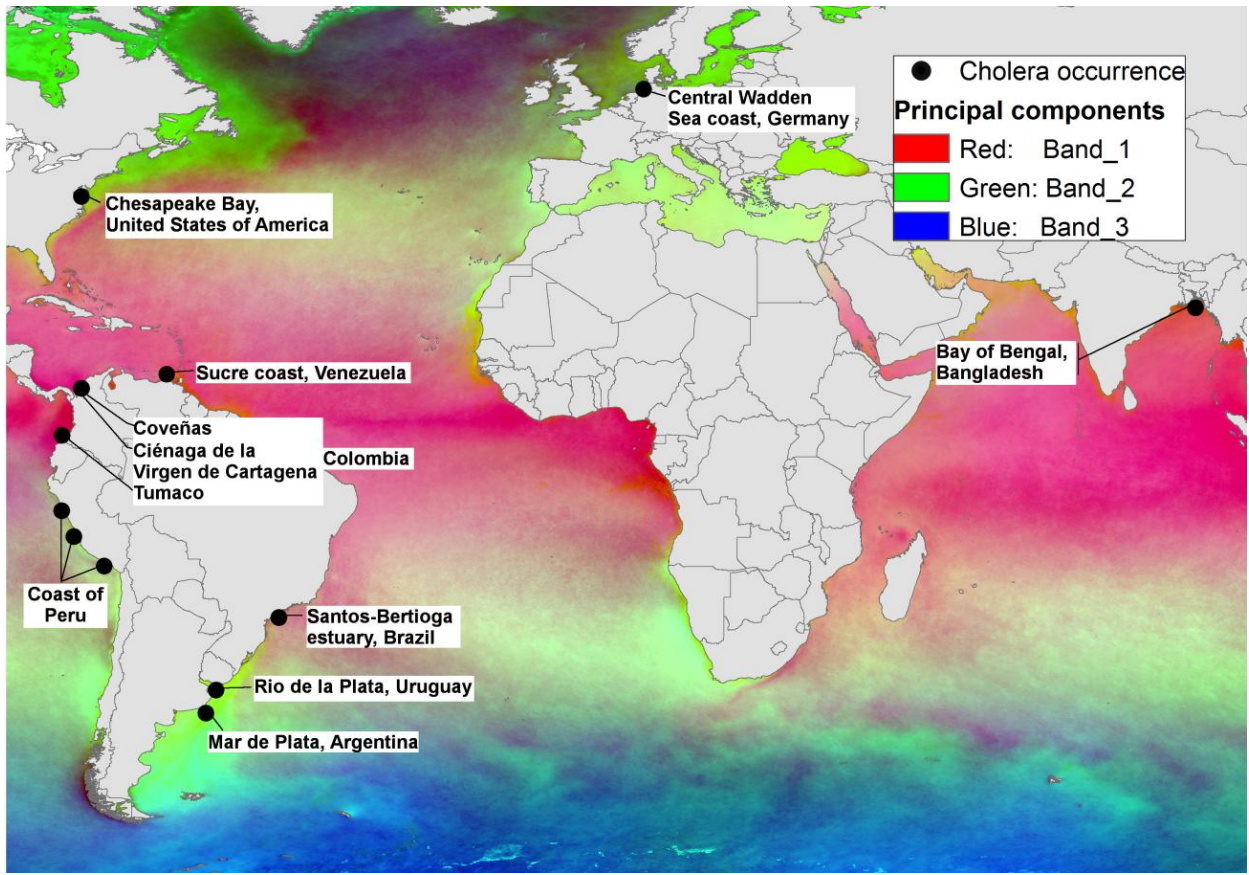

**Figure 1. Reports of *Vibrio cholerae* from sea water samples.** Sites of *V. cholerae* presence (black dots) represent diverse geographic areas and marine environments. *V. cholerae* presence was observed in United States (Louis et al., 2003), Venezuela (Grau et al., 2004), Colombia (López et al., 2010), Peru (Lipp et al., 2003; Orozco et al., 1996), Brazil (Martinelli Filho et al., 2010), Uruguay (Binsztein et al., 2004), Argentina (Binsztein et al., 2004), Germany (Böer et al., 2013), and Bangladesh (Mukhopadhyay et al., 1998; Pal et al., 2006; Pascual et al., 2000; Sack et al., 2003). Colors represent environmental principal components used to measure environmental similarity between sites of *V. cholerae* occurrence: one (red), two (green), and three (blue) with similar colors representing similar environmental conditions.

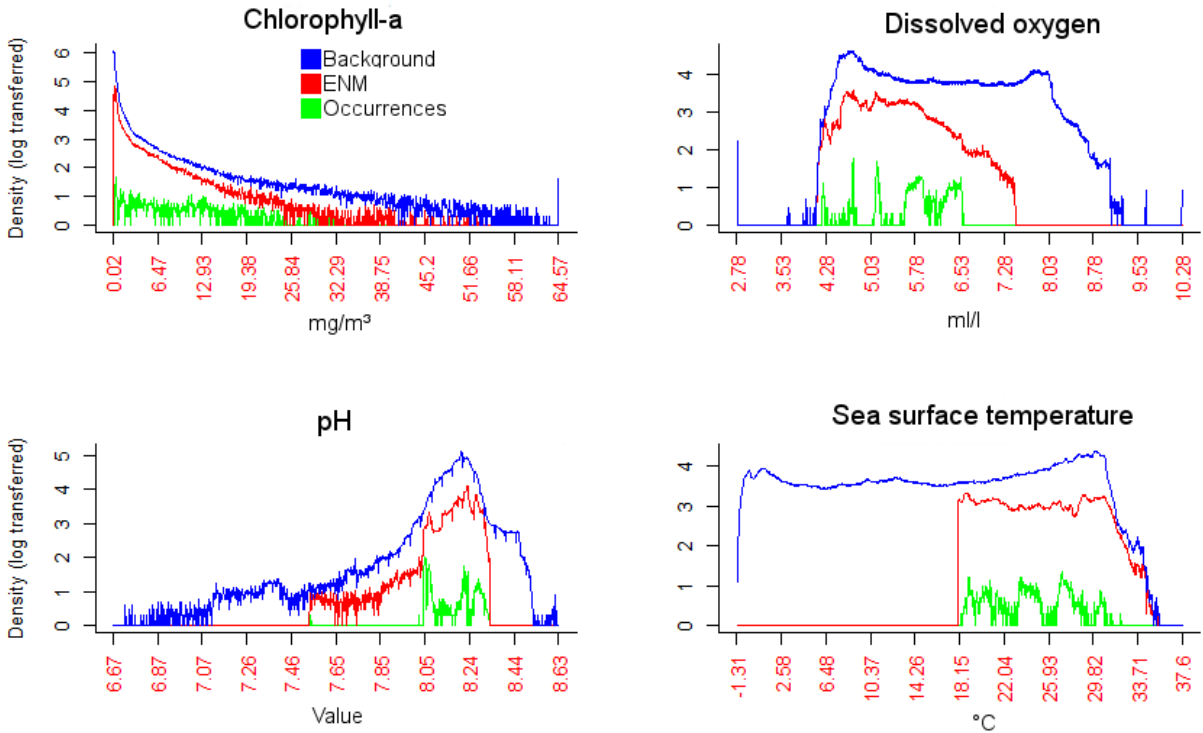

**Figure 2. Global distribution of *Vibrio cholerae* across key environmental parameters from the 15 sites.** On the X axis, mean chlorophyll-a (top left), pH (bottom left), concentration of dissolved oxygen (top right), and maximum sea surface temperature (bottom right) are presented, and on the Y axis is the number of pixels for each variable. Available values in the oceans (blue) are plotted to demonstrate the full environment parameter space and the ecological niche occupied by *V. cholerae*, according to the ecological niche model (ENM; red) and the sites of laboratory-confirmed *V. cholerae* presence (green).

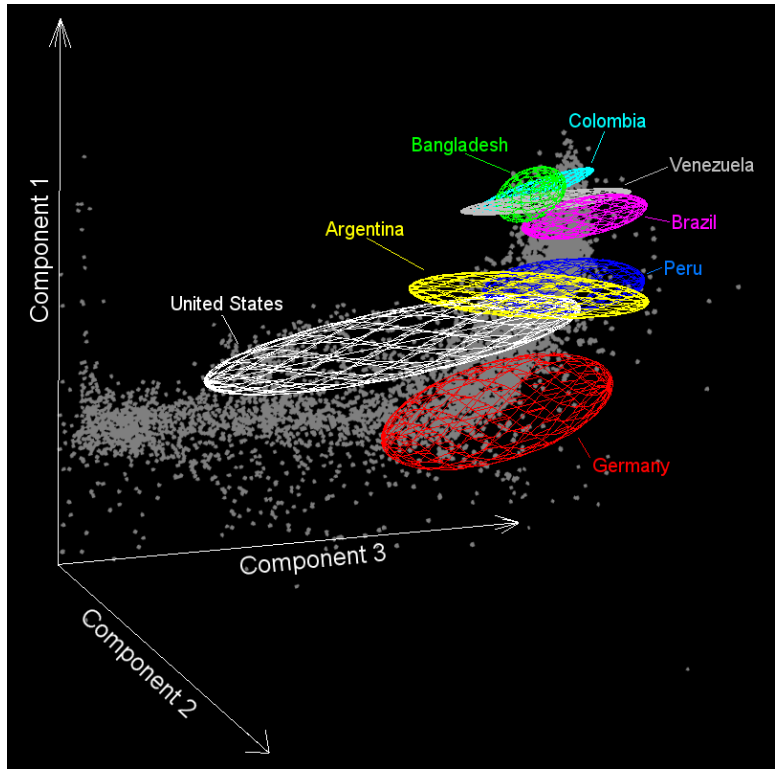

**Figure 3. Ellipsoids of *Vibrio cholerae* presence into the environmental space.** Axes of the environmental space are principal components 1, 2, and 3 from the oceanographic variables (see Methods). Overlapping ellipsoids indicate similar marine conditions. Bangladesh, a region where cholera is endemic, is environmentally similar to areas in Latin America (e.g., Colombia, Venezuela, Brazil). The United States demonstrates a broad range of environments available for the potential distribution of *V. cholerae*, although the geographic area included in the model was small (i.e., Chesapeake Bay).

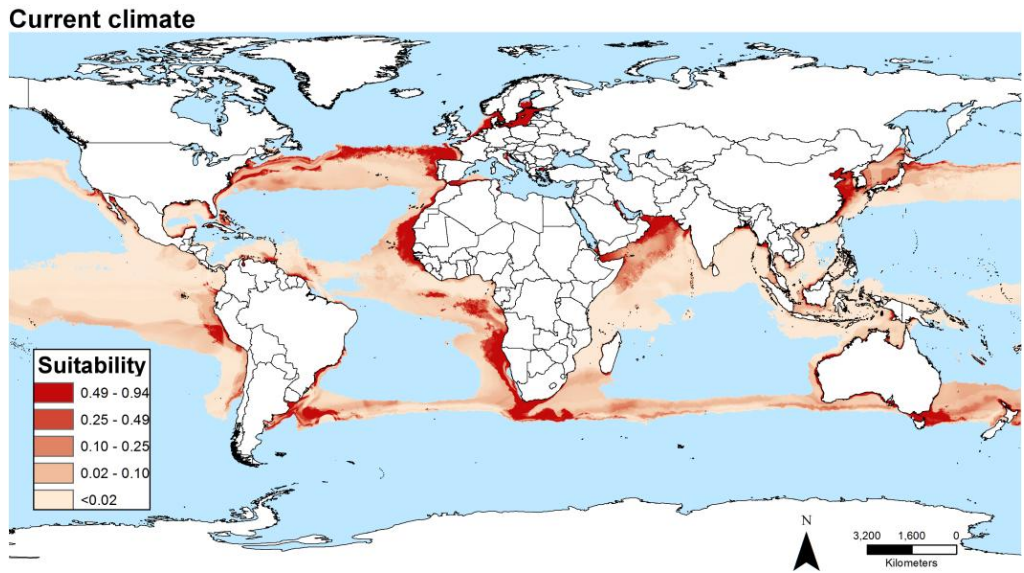

**Figure 4. Global map of *Vibrio cholerae* suitability in seawaters.** Model generated under the current climate scenario.

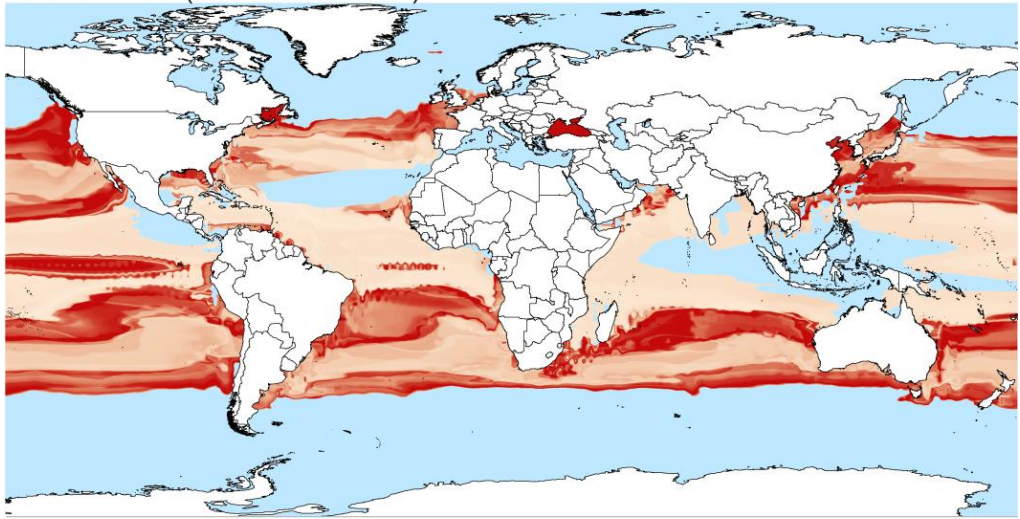

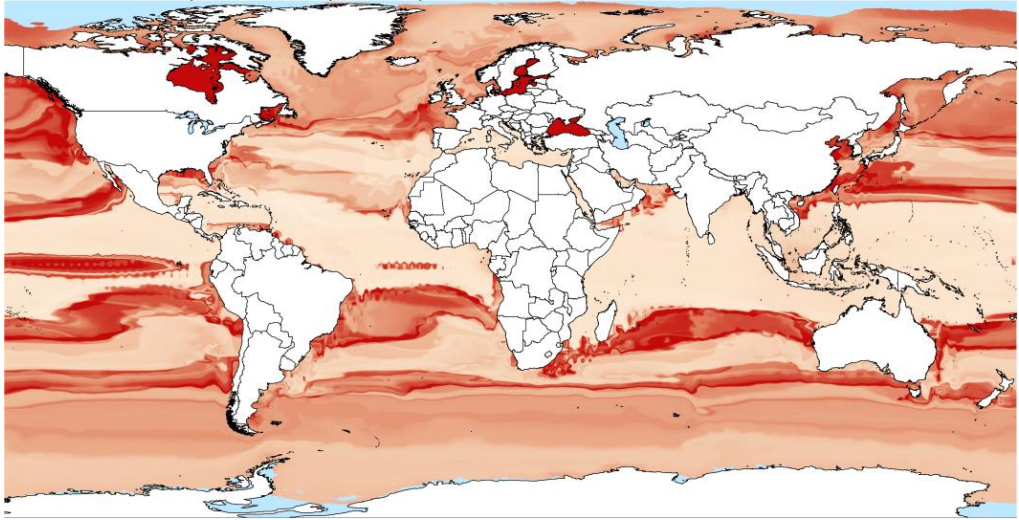

**Figure 5. Global map of *Vibrio cholerae* suitability in seawaters under future B1 (IPCC) climate scenarios.** Models calibrated with current sea surface temperature and salinity were transferred (top) and extrapolated (bottom) to values for the year 2100.

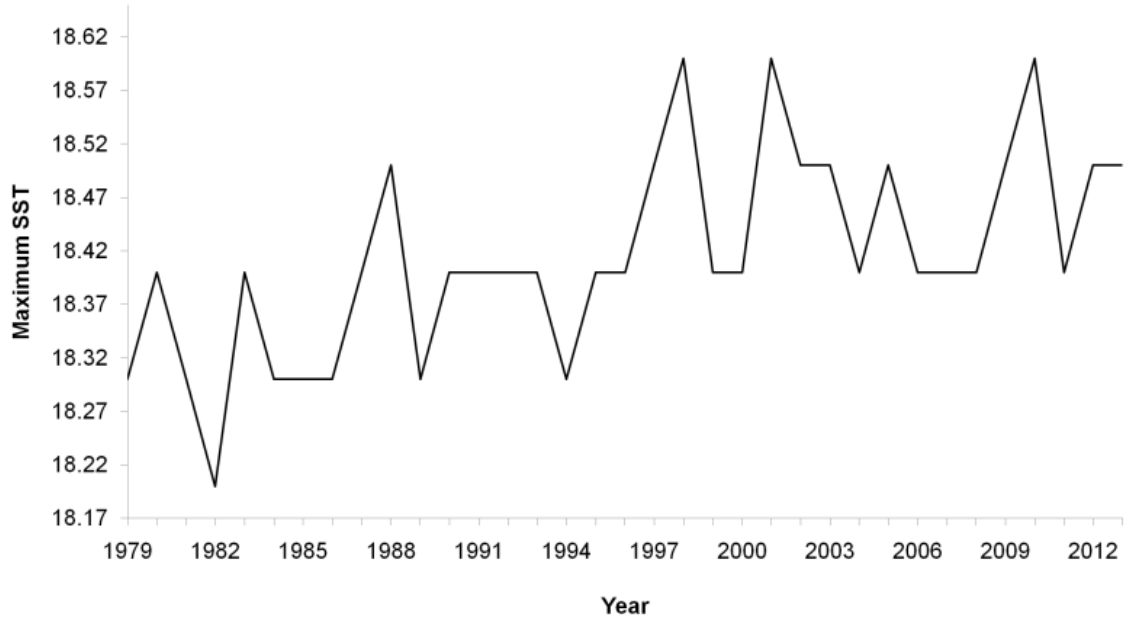

**Supplementary Material S1. Global maximum Sea Surface Temperature (SST; °C).** The maximum month SST value is presented by year. Average monthly SST values from January, 1979 to December, 2013 are available at http://cci-reanalyzer.org

**Supplementary Material S2. Correlation matrix for the 12 oceanographic variables included in the principal component analysis.**

| Variable | Chlomax | **Chlomean** | Cholmin | Dissox | Nitrate | Parmax | **pH** | Phos | **Salinity** | **SSTmax** | SSTmean | SSTmin |
|---|---|---|---|---|---|---|---|---|---|---|---|---|
| Chlomax | 1 | | | | | | | | | | | |
| **Chlomean** | 0.923 | 1 | | | | | | | | | | |
| Cholmin | 0.74 | 0.92 | 1 | | | | | | | | | |
| Dissox | 0.13 | 0.12 | 0.09 | 1 | | | | | | | | |
| Nitrate | -0.01 | -0.01 | -0.02 | 0.85 | 1 | | | | | | | |
| Parmax | -0.08 | -0.07 | -0.049 | -0.81 | -0.75 | 1 | | | | | | |
| **pH** | -0.23 | -0.25 | -0.21 | -0.35 | -0.32 | 0.35 | 1 | | | | | |
| Phos | 0.01 | 0.01 | -0.01 | 0.87 | 0.95 | -0.77 | -0.36 | 1 | | | | |
| **Salinity** | -0.49 | -0.52 | -0.43 | -0.40 | -0.17 | 0.35 | 0.20 | -0.23 | 1 | | | |
| **SSTmax** | -0.06 | -0.06 | -0.04 | -0.97 | -0.90 | 0.80 | 0.35 | -0.90 | 0.29 | 1 | | |
| SSTmean | -0.11 | -0.11 | -0.07 | -0.99 | -0.86 | 0.79 | 0.34 | -0.87 | 0.35 | 0.98 | 1 | |
| SSTmin | -0.15 | -0.14 | -0.10 | -0.98 | -0.82 | 0.76 | 0.33 | -0.82 | 0.37 | 0.96 | 0.99 | 1 |

Bolded: variables included in the ecological niche model.

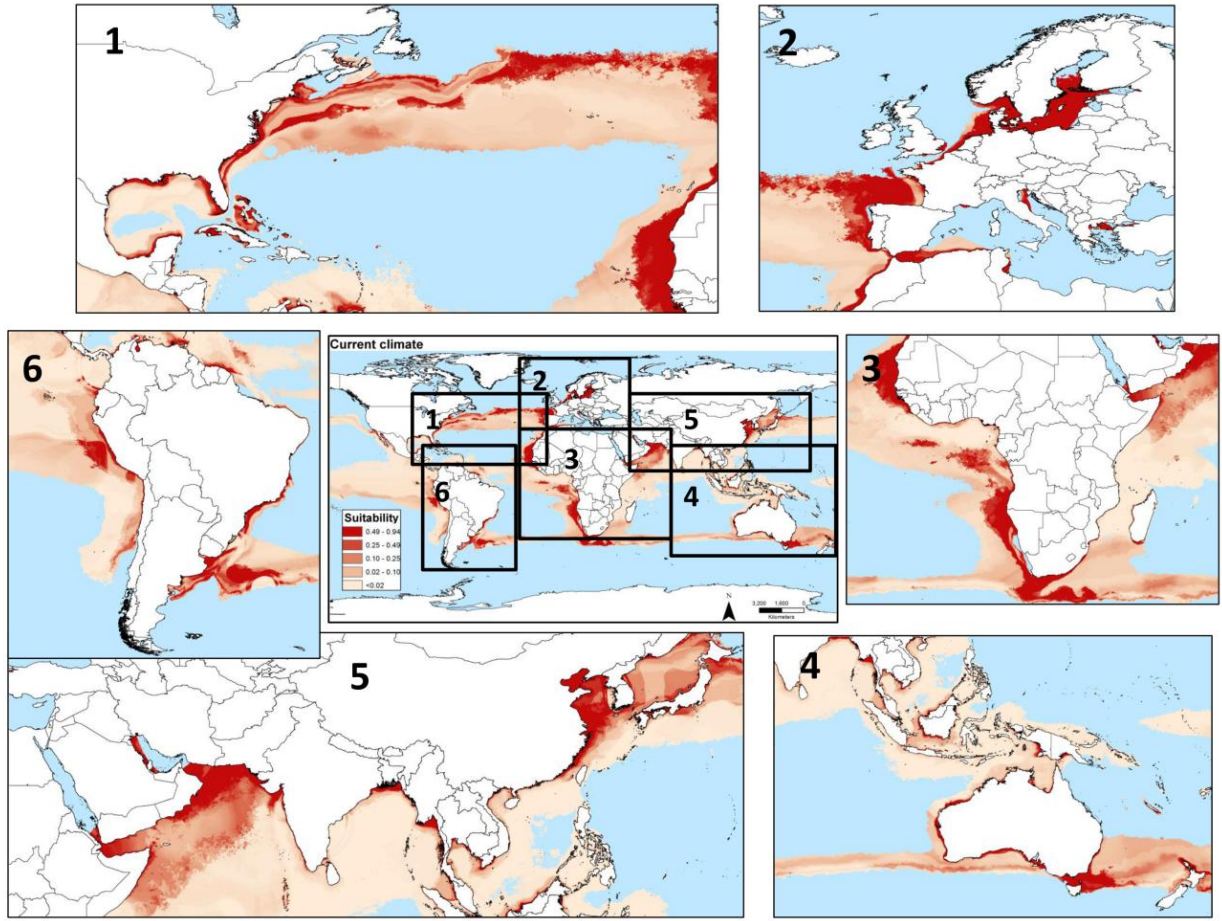

**Supplementary Material S3. Foci of high probability regions.** 1. Eastern coastal North America, parts of Central America, and the Atlantic; 2. Europe, particularly highlighting coastal Spain, and Scandinavian waters; 3. Africa and portions of the Middle East; 4. Oceania and portions of Southeast Asia; 5. Asia, highlighting coastlines from the Middle East to Japan; and 6. South America, noting high probability areas on both East and West coastlines.

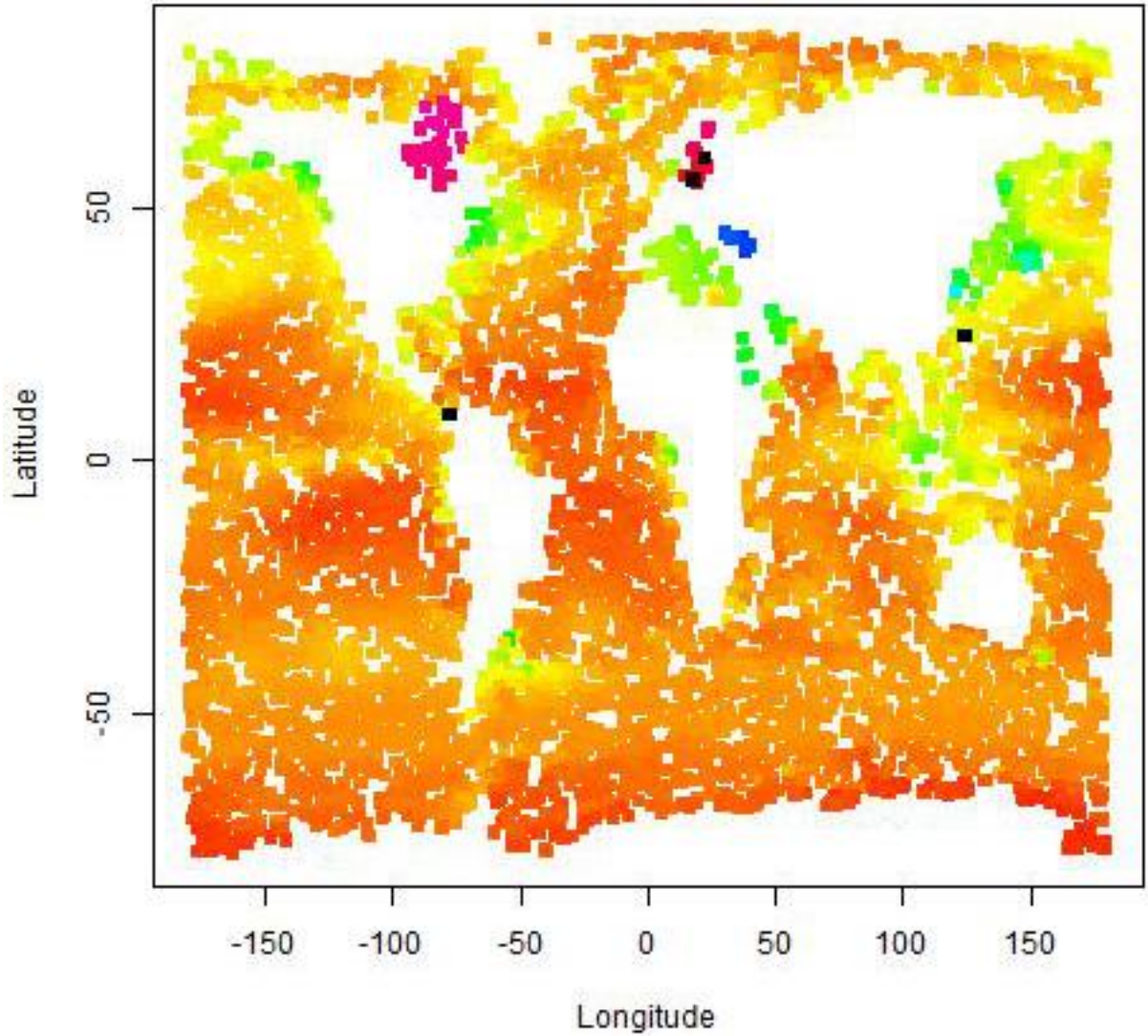

**Supplementary material S4. Identification of future marine environmental conditions not available at present and marine environmental conditions similar between current and future climate scenarios.** The Mobility-Oriented Parity (MOP) analysis indentifies areas where model should not forecast due to the absence of such environmental conditions under the current climate scenario (black areas). The Multivariate Environmental Similarity Surface (MESS) analysis identifies areas with highly (red) or moderately (blue) similar environmental conditions between current and future climate scenarios.